# Designing Perturbative Metamaterials from Discrete Models: From Veselago lenses to topological insulators

Kathryn H. Matlack*[1,3], Marc Serra-Garcia*[1a], Antonio Palermo[2], Sebastian D. Huber[3], Chiara Daraio[1,4]

[1]*Department of Mechanical and Process Engineering, ETH Zürich, Zürich, Switzerland*

[2]*Department of Civil, Chemical, Environmental and Materials Engineering - DICAM, University of Bologna, Bologna, Italy*

[3]*Institute for Theoretical Physics, ETH Zürich, Zürich, Switzerland*

[4]*California Institute of Technology, Pasadena, CA*

**\***These authors contributed equally to this work

a) sermarc@ethz.ch

**Abstract**

Discrete models provide concise descriptions of complex physical phenomena, such as negative refraction, topological insulators, and Anderson localization. While there are multiple tools to obtain discrete models that demonstrate particular phenomena, it remains a challenge to find metamaterial designs that replicate the behavior of desired nontrivial discrete models. Here we solve this problem by introducing a new class of metamaterial, which we term "perturbative metamaterial", consisting of weakly interacting unit cells. The weak interaction allows us to associate each element of the discrete model (individual masses and springs) to individual geometric features of the metamaterial, thereby enabling a systematic design process. We demonstrate our approach by designing 2D mechanical metamaterials that realize Veselago lenses, zero-dispersion bands, and topological insulators. While our selected examples are within the mechanical domain, the same design principle can be applied to acoustic, thermal, and photonic metamaterials composed of weakly interacting unit cells.

**Introduction**

Metamaterials utilize sub-wavelength structures to control wave propagation, achieving extreme functionalities such as focusing beyond the diffraction limit[1–5], performing mathematical operations with light[6], or cloaking objects[7,8]. While the potential of metamaterials is well established, we lack systematic approaches to metamaterial discovery, which is currently based on intuition, trial and error, or unguided searches of large design spaces. Data-driven design approaches[9–11] have been successful in engineering quasi-static material performance, but extending them to dynamics requires taking into account the interactions between multiple vibrational modes of the unit cell. A promising approach towards metamaterial design in the dynamic regime draws inspiration from the field of electronics[12–14], where complex devices are designed by combining discrete "lumped" elements, such as capacitors or inductors. These lumped elements are able to capture the intended behavior independently of the implementation details of the elements, which greatly simplifies the design process. Once a phenomenon has been described as a lumped element model, it can be easily implemented in different domains. In the mechanical domain the capacitors and inductors can be replaced by masses and springs, and implemented in platforms as diverse as pendulum arrays[15] or networks of piezoelectric resonators[16]. While it is possible to implement lumped models by combining discrete masses, springs, capacitors or inductors, there is currently no systematic design process to convert these models to metamaterials. This is because the space of possible designs is extremely large, and the relationship between a structure and its dynamical properties is highly non-trivial.

We propose a generic tool to solve this inverse problem: we map building blocks of a metamaterial to components of a mass-spring model, and determine the metamaterial design by searching the space of possible combinations of building blocks. In order for this approach to be successful, we need to address two key challenges. First, we need an efficient algorithm to extract a reduced order model from the metamaterial design. Second, we need a system where changes to different parts of the design do not interfere with each other, such that the design elements are additive in their effect. This additive property allows us to divide the search space into much smaller independent sub-spaces, resulting in an exponential speedup of the search process.

In this paper we successfully address these two challenges with "perturbative metamaterials": systems consisting of unit cells with a spectrum of linear normal modes that weakly interact with modes of neighboring unit cells. We obtain reduced-order models for our metamaterial through a method adapted from quantum material science: we use the Schrieffer-Wolff transformation[17,18] to isolate modes in the frequency range of interest. The purpose of extracting reduced order models is to efficiently quantify how design changes affect the material's dynamic properties. We then catalogue these changes for various geometries, and optimize the configuration of the metamaterial's components to obtain the target mass-spring model. We show that with a suitable series expansion of the Schrieffer-Wolff transformation, we can explore a space containing on the order of $10^{30}$ design configurations, which is impossible to do using optimization methods with current computational power. We demonstrate this in a system of plates connected by soft beams, which achieves the additive property essential for an effective algorithm.

We show the potential of our generic scheme on three key examples: a Veselago lens, a zero group velocity material, and a topological insulator, each with increasing complexity in their unit cell designs.

**Extracting a reduced order model from a perturbative metamaterial**

Here we present a method to extract a reduced order model for a given metamaterial design (e.g., Fig. 1a), which is a pre-requisite for our design procedure. The method works for arbitrary unit cells as long as they interact weakly. We start by selecting the frequency range where the model will be valid. Each unit cell mode that falls within that range (Fig. 1b) translates to a degree of freedom in the reduced-order model. Our reduced order model calculation (Fig. 1a-c) differs from existing reduction techniques[19–25] in that it constructs an approximation of a complex metamaterial geometry by combining results obtained by considering geometric features individually. This property is crucial when designing metamaterials from discrete models. We explain our method using a system of square lattice of steel plates (10mm x 10mm x 0.5mm) weakly interacting through soft, polymer beams. The plates can also contain holes that tune the local resonance frequencies. We emphasize that this system is chosen as an example, and our method can be applied to any metamaterial that satisfies the weak interaction condition.

The reduced-order models are based on a local modal expansion and consist of local resonators coupled through springs. Each local resonator corresponds to a plate mode that falls within our frequency range of interest, and the springs account for the coupling introduced by the beams. We refer to a system of plates without beams as "uncoupled", and a system of plates connected with beams as "coupled" (Fig. 1b,c). For clarity, we define two separate coupling terms: $V$ is the coupling between modes in the full physical system, $V^R$ is the coupling in the reduced-order model that we extract.

Assuming the beams are short enough such that we can consider their effect as instantaneous and neglect their internal degrees of freedom (DOFs), the equation of motion of the coupled system is:

$$\ddot{u} + ([M + \Delta M]^{-1}[K + \Delta K])u = 0, \tag{1}$$

where $u$ is the displacement, $M$ and $K$ represent the effective mass and effective stiffness of the plates, and $\Delta M$ and $\Delta K$ represent the contributions of the beams. Since we have weakly interacting unit cells, we can neglect higher powers of $\Delta K$ and $\Delta M$, such that:

$$\ddot{u} + (H_0 + V)u = 0. \qquad (2)$$

Here, $H_0 = M^{-1}K$ is the diagonalized dynamical matrix of the uncoupled system, $(H_0 + V)$ the dynamical matrix of the coupled system, and $V = M^{-1}(\Delta K) - M^{-1}(\Delta M)M^{-1}K$ a perturbation on the uncoupled system. In general, the perturbation $V$ couples each mode of the unit cell to each mode of neighboring unit cells. This prevents us from restricting our description to modes that only lie in our frequency range of interest, i.e. $H_0 + V$ is not block-diagonal, but instead contains coupling terms between our modes of interest and irrelevant modes. To remove the coupling between relevant and irrelevant spaces, we perform a suitable rotation, called the Schrieffer-Wolff (SW) transformation, of the dynamical matrix. Originally developed in the context of the Anderson model of magnetic impurities in metals[17], the SW transformation is the rotation matrix $U$ such that $U(H_0 + V)U^T$ is block diagonal[18]. This model reduction is a staple in quantum information and condensed matter physics[26,27]; here we use it to analyze dynamics of mechanical metamaterials.

The SW transformation can be calculated perturbatively, with the expansion parameter[18]

$$\epsilon = \frac{V_{ij}}{E_i - E_j}, \tag{3}$$

where $E_i$ are $E_j$ are eigenvalues of the uncoupled system, and $V_{ij}$ is the coupling between modes *i* and *j*. This expansion parameter can be interpreted as the strength of the coupling relative to the spectral gap between the mode of interest and other modes. For small coupling values, the first-order term provides a satisfactory approximation (Fig 1e). This first-order term is linear, which means the contributions of the individual geometric elements (beams and holes in the plate system) are additive, which is of crucial importance for our algorithm. Higher orders of the SW transformation provide a more accurate reduced-order description of the system in the presence of stronger couplings (Fig 1f, g), but contain long-range interactions, i.e., stiffness terms that couple plates not physically connected by beams (*Supplementary Information*).

In this work, we extract the reduced-order coupling matrix $V_{ij}^R$ from a finite element (FE) simulation (*Methods*). To determine the effect of the beams, we first calculate a large number of modes for a system of two coupled plates (Fig 1b). Then, we express these coupled modes as a linear combination of uncoupled plate modes, by using a least-squares approximation of the plate's displacement within a test area (Fig 1c). We use this representation to determine the dynamical matrix of the coupled system, by inverting the eigenvalue problem, $D = P^{-1}(H_0 + V)P$. Here, $D$ is the matrix whose diagonal contains the eigenvalues of the coupled system and $P$ is a matrix whose columns contain the modal displacements of the coupled system expressed as a linear combination of uncoupled plate modes. We determine the perturbation introduced by the holes by following an analogous procedure: We calculate the modes for a plate with holes, express these as a linear combination of unperturbed plate modes and invert the eigenvalue problem. By summing the contribution of the beams and holes, we determine $V_{ij}$, which is finally SW-transformed resulting in the reduced-order model $V_{ij}^R$.

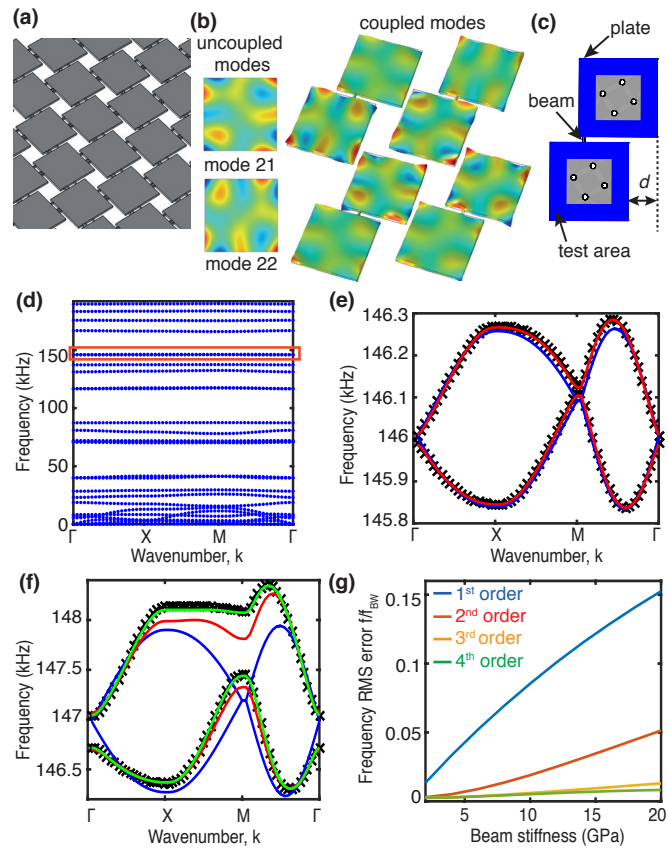

**Figure 1 | Method of extracting a reduced-order model. a.** Test material, made of steel plates connected with polymer beams. **b.** Uncoupled degenerate plate modes (left) and modes of the

coupled plate system (right). **c.** Two plates coupled with one beam and multiple holes to adjust the plate's local stiffness. The displacement is sampled in the blue area and used for the coupling stiffness calculation. **d.** Example 2D band structure for a plate-beam metamaterial and frequency range of interest, around 145 kHz encompassing 2 plate modes. Dispersion relation for FE simulation (black data points) and reduced order model for: **e.** system with soft couplings (beams with Young's modulus of 4 GPa) compared with first and second order SW transformation, and **f.** system with larger coupling strength (beams with Young's modulus of 20 GPa) compared with first, second, and fourth order SW transformations. **g.** Error in the frequency relative to the bandwidth frequency ($f_{BW}$) predicted by the SW transformation, for increasing coupling strengths.

**Metamaterial design from a discrete model**

We apply the reduced-order modeling method to design a metamaterial, by finding the optimal configuration of plates, beams, and holes to match the target mass-spring model, whose coupling matrix is denoted $V^T$. We limit the beam stiffness to small values, such that the first order SW transform provides a good first approximation. As a consequence, we can assume changes in our design are additive in their effect (*Supplementary Information*). This means we can easily explore a large space of possible configurations (containing up to $10^{30}$ elements), by optimizing different parts of the design separately (each design subspace containing less than $10^{10}$ elements), and by adding combinations of calculated responses from a few FE simulations ($\sim 10^2$) of different beam and hole parameters. The design examples illustrate this concept.

The first step of our method is to map the discrete model's DOFs into plate modes, taking into account that a single plate can map to multiple DOFs by utilizing multiple plate modes, e.g., degenerate modes (as in the "zero group velocity material" and "topological insulator" design examples). This mapping is done manually, and there are multiple acceptable ways to accomplish it.

The second step is to introduce the desired inter-plate couplings, by determining the beam parameters: location and thickness, and in some cases the angle (e.g., in the "topological insulator" example). To do this, we pre-compute a table of coupling stiffnesses for different beam locations and plate alignments ($d$ in Fig. 1c). We then perform a combinatorial search, exploiting the additivity of the first-order SW transformation, to find a set of beams that add up to the desired coupling (*Methods*). Finally, we perform a gradient-based optimization to compensate for second-order effects due to the finite beam stiffness. This optimization is seeded by the approximate solution from the combinatorial search and is done on a test system containing two plates (Fig. 1c).

The third step is to tune each plate's local response, i.e., each mode's frequency and the couplings between different modes of the same plate. This tuning corrects for the local effect of the beams, which shift the plate frequencies and introduce couplings between different modes of the same plate. We first add up the local contribution of all beams connected to a plate and then determine the necessary adjustment to match the desired local properties. This adjustment is obtained by introducing holes in the plate. Since the plate modes have different displacement profiles, holes in different locations will have different effects in each of the modes. We determine the hole radii and locations by the same procedure that we used with the beams: (1) create a table of the hole's effect in different locations, (2) perform a combinatorial search on the hole table, and (3) perform a gradient-based optimization on the results of the combinatorial search.

As a fourth and final step, we perform a final gradient-based optimization on a system containing multiple unit cells. This optimization includes all holes and beams, and therefore is able to compensate for second-order effects that arise due to the interactions of beams and holes as well as long-range couplings (*Supplementary Information*).

**Design examples**

*Phononic Veselago lens*

We first design a classical Veselago lens[28] metamaterial as an example to demonstrate how our design method works. We choose this example because the Veselago lens is a well-understood system that has been demonstrated in optical[1] and phononic[5] systems.

In the phononic Veselago lens, a double negative medium, i.e., a medium with negative effective modulus (-$K$) and effective density, is embedded in a conventional medium of equal but positive modulus (+$K$) and effective density (Fig. 2a). We create a mass-spring system that approximates the lens in a square lattice. The basic unit cell consists of a single resonator connected by springs to its 4 nearest neighbors (Fig. 2b).

The target coupling stiffness between unit cells is either +K or –K. The effective masses are obtained by shifting the local resonance frequencies of the sites inside and outside the lens region. The normalized effective mass of a harmonic oscillator at a particular frequency is $M_{eff}(\omega) = \left(1 - \frac{\omega_0^2}{\omega^2}\right)$, where $\omega_0$ is the resonance frequency of the mode and $\omega$ is the lens' frequency of operation. The Veselago lens requires $M_{lens}(\omega) = -M_{medium}(\omega)$. The relation between the local resonance frequency of the sites inside and outside the lens region is thus:

$$\omega_{lens}^2 = 2\omega^2 - \omega_{medium}^2. \qquad (4)$$

We choose the plate mode 24 to couple between unit cells, which has a high enough stiffness compared to the beams, and is well separated from neighboring modes. Since the mode profiles at the boundaries have 90° rotational symmetry, if the plates are perfectly aligned, couplings of only one sign are possible. To overcome this limitation, we offset the plates so we can find beam positions to design both positive and negative couplings (Fig. 2c). We choose the beam parameters that give a positive or negative coupling based on the pre-computed coupling stiffnesses (Fig. 2d). We then choose the hole parameters that compensate for the local stiffness introduced by the beams, such that the unit cells inside and outside the lens have the required local resonance frequencies from equation (4) (Fig. 2e).

We analyze the resulting metamaterial lens using FE simulations (*Methods*). The results clearly illustrate the Veselago lens effect (Fig. 2f), and show excellent agreement with the results of the mass-spring model (Fig. 2g).

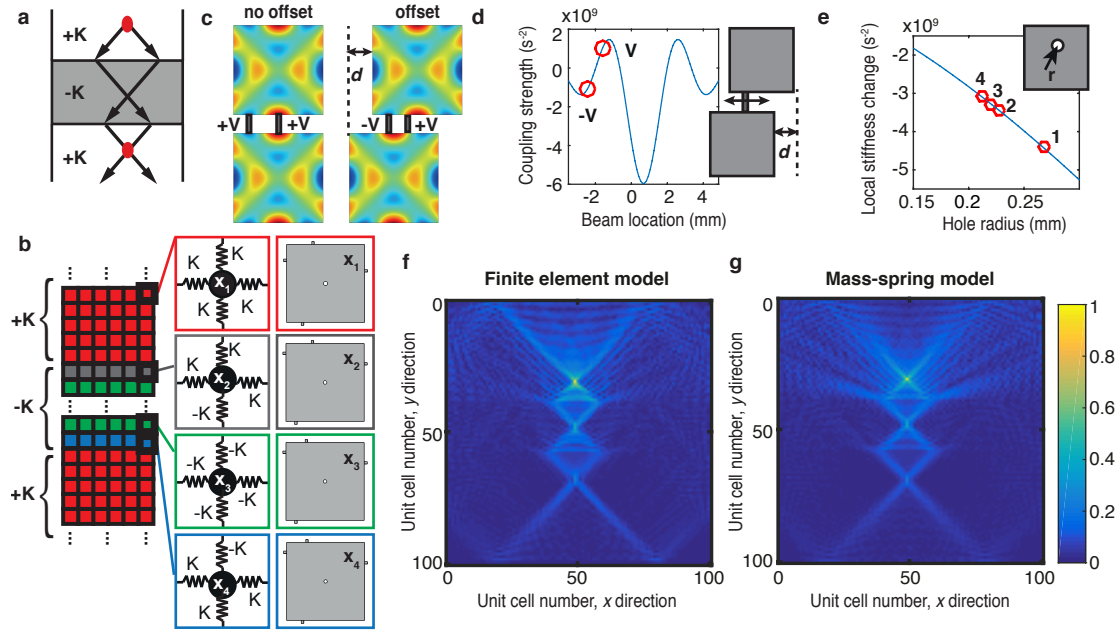

**Figure 2 | Veselago lens metamaterial example. a.** A schematic of focusing in the Veselago lens. **b.** Mass-spring model of the lens, showing construction of the four different unit cells and their corresponding metamaterial design: a unit cell with all positive springs (red), a unit cell with all negative springs (green), and two interface unit cells with a combination of positive and negative springs (gray and blue). Lattice sites ($x_1$, $x_2$, $x_3$, $x_4$) are local resonators with their own mass and spring. The mass-spring model contains 100x100 unit cells, where the double negative region consists of nineteen rows in the center. **c.** Illustration of plate offset concept with mode 24: when plates are aligned (left), only couplings of one sign are possible. If plates are offset by distance *d,* both

positive and negative couplings are possible. **d.** Calculated coupling stiffness for different beam locations at a given offset, where data points show locations of the beams to achieve positive and negatives stiffness. **e.** Calculated local stiffness change for different hole radii, where data points show the radii in each of the four unit cells for the intra-plate coupling compensations. **f.** Results of metamaterial lens from FE simulations, at 175.284 kHz. **g.** Results of mass-spring model lens, at 175.204 kHz. The color bar applies to both **f** and **g,** and indicates the normalized amplitude of the RMS displacement.

*Zero group velocity material*

We now design a zero group velocity ($c_g$) material, which has a flat band within a lattice with 1D periodicity, as a way to demonstrate the use of multiple and degenerate modes of a single plate in our design method. The zero $c_g$ material is analogous to a quasi-1D Lieb lattice, and is of interest as it is a perfectly periodic configuration that leads to a flat band without using defects. Such localized states have been explored recently in photonic waveguides[29,30], and could have applications for slow phonon modes in elastic metamaterials.

In this example, the target unit cell consists of three equal resonators in an "L" configuration (Fig. 3a). We start by mapping every DOF ($x_1$, $x_2$, $y_1$) to a plate mode. We choose a unit cell design consisting of two plates (P1 and P2). We map $x_1$ and $y_1$ to a pair of degenerate plate modes (21, 22) in P1, and $x_2$ to one mode in P2, within the same degenerate pair. This leaves us with an extra mode within our frequency range of interest, $y_2$ (Fig. 3b), which we remove from the dynamics by shifting its frequency outside the range of interest, through the introduction of holes in the plate. We point out that this mode assignment is not unique and we could, e.g., use non-degenerate modes and a three-plate unit cell design. We choose to use degenerate modes to illustrate how we can exploit them, which is crucial in the next design example, the topological insulator.

The equation of motion of the system in Fig. 3b is: $\ddot{u}_i = V^T_{i\mathrm{mod}2}\, u_i + V^T_m u_{(i+1)} + V^T_m u_{(i-1)}$. The vector $u$ contains the two DOFs $x$ and $y$, and $i$ is the index of the unit cell. The target coupling matrices for the inter-plate couplings ($V^T_m$), and the intra-couplings ($V^T_0$, $V^T_1$) are:

$$V^T_m = \begin{bmatrix} \alpha & 0 \\ 0 & 0 \end{bmatrix},\ V^T_1 = \begin{bmatrix} \beta & \alpha \\ \alpha & \beta \end{bmatrix},\ V^T_0 = \begin{bmatrix} \beta & 0 \\ 0 & \beta + \Delta \end{bmatrix} \tag{5}$$

where $\alpha$ is the coupling strength, $\beta$ is the local stiffness of the plate, $\alpha \ll \beta$, and $\Delta$ is the compensation for the extra mode.

In this metamaterial design, we illustrate the concept of using one plate to implement multiple DOFs (Fig. 3b,c). We chose the degenerate plate modes 21 and 22 (Fig. 1b), which is motivated by the following requirements: (1) there is good separation between the degenerate pair of interest and neighboring modes; and (2) the mode profiles at the boundary exhibit a complex structure that enable a wide variety of coupling stiffness.

We do a combinatorial search to determine the plate offset, and thickness and position of the beams, finding that three-beam combinations can satisfy the target coupling stiffness matrix $V_m^T$. We then find the plate hole locations to satisfy the target matrices $V_1^T$ and $V_0^T$. We perform an additional gradient-based optimization on multiple unit cells to reduce the second order error between the metamaterial and target stiffness (*Supplementary Information*).

The dispersion relation for the zero $c_g$ metamaterial compared to the mass-spring model is shown in Figure 3d. The metamaterial dispersion shows excellent correspondence to the mass-spring system. While there is some slight non-zero group velocity in the flat band from long-range couplings (*Supplementary Information*), our metamaterial approaches the desired the mass-spring model with very good precision.

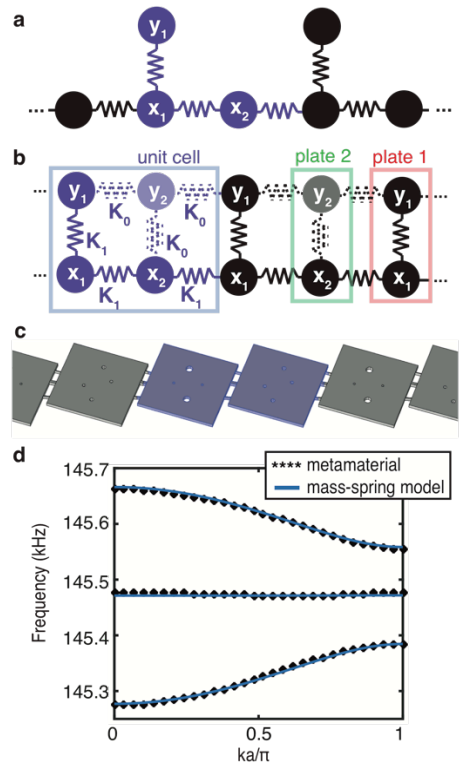

**Figure 3 | Zero group velocity metamaterial. a.** Zero $c_g$ material mass-spring model. **b.** Mass-spring model used for metamaterial design, where each vertical column of two masses $x$ and $y$ represents two degenerate modes of one plate in the metamaterial. The metamaterial is optimized to separate mode $y_2$ in plate 2 from the other modes, and to push $K_0$ to 0. **c.** Design of the zero $c_g$ metamaterial, with unit cell highlighted in purple. **d.** Dispersion curves for the zero $c_g$ material mass-spring model compared to the designed metamaterial. Only the three energy bands of interest are shown for clarity.

*Topological insulator*

Topological insulators are a unique class of materials that are electrically insulating in the bulk, yet conductive on the surface, and whose surface states are immune to back-scattering and defects[31–33]. These materials have inspired a new class of mechanical systems that mimic these topological insulator properties in the elastic domain[15,16,34–38]. Researchers have developed a mapping between topological spin-orbit systems and discrete mechanical lattices[39,40], which are highly complex. Here, we focus on the mass-spring model proposed in ref. 15, which contains 6 DOFs per unit cell that interact nontrivially with neighboring unit cells. While this mass-spring model was realized with a discrete system of pendula[15], the implementation of such models in metamaterials is an open research problem, which our method is able to solve.

The mass-spring model unit cell for this topological insulator consists of three 2-DOFs lattice sites[15] (Fig. 4a). The equation of motion is $\ddot{u}_{ij} = V_L^T u_{ij} + V_0^T u_{(i+1)j} + V_0^T u_{(i-1)j} + V_{i \bmod 3}^T u_{i(j+1)} + V_{i \bmod 3}^T u_{i(j-1)}$. The vector $u$ contains the two DOFs $x$ and $y$, and $i$ and $j$ are the row and column indices of the unit cells in Fig. 4a. The target inter-plate coupling matrices are:

$$V_n^T = \alpha \begin{bmatrix} \cos(2\pi n/3) & \sin(2\pi n/3) \\ -\sin(2\pi n/3) & \cos(2\pi n/3) \end{bmatrix} \tag{6}$$

where $n$ is an integer that spans from 0 to 2. The intra-plate coupling matrix is $V_L^T = I_2 \beta$, where $I_2$ is the identity matrix, and $\beta$ and $\alpha$ are as defined above.

The designed metamaterial translates each 2-DOF site into a single plate, by using the degenerate plate modes 21 and 22. Thus the unit cell consists of 3 plates coupled with beams (Fig. 4b), which are optimized to match the required $V_0^T, V_1^T$, and $V_2^T$ matrices. By performing separate combinatorial optimizations for each coupling, we reduce the problem to three searches on a space of $10^{10}$ configurations instead of a single search on a space of $10^{30}$ configuration. Our method cannot obtain a solution for all three couplings at a single offset, so we introduce a third beam parameter: an angle. A non-zero beam angle allows for different plate alignments using the same offset for the three different couplings (Fig. 4b).

We calculate the band structure of the metamaterial with 1D periodicity (*Methods*). The dispersion relation (Fig. 4c) shows the expected three bands of bulk modes, separated by two counter-propagating edge modes that cross at $\pi/3$ and $2\pi/3$. We then model a finite 7x7 plate system and calculate the eigenmodes with a FE simulation, and the results clearly show edge modes (Fig. 4d). The metamaterial shows excellent agreement to the target mass-spring model (*Supplementary Information*). The designed metamaterial also illustrates the stability of topologically protected modes against defects. We explore this by introducing a defect in the metamaterial, which we model as three fixed plates. The edge mode still persists, and propagates around the defect (Fig. 4e).

Micro-fabrication of these composite metamaterials should result in high-speed devices with advanced signal processing capabilities, and benefit engineering applications in communication systems, ultrasonic imaging, and filtering for surface acoustic waves. Further exploration should aim to increase the bandwidth and reduce constraints in the material parameters, as well as to deliberately introduce nonlinearity. The same design method that we have presented here can be applied to electromagnetic systems[12,41], enabling the design of arbitrary photonic circuits with extreme computing capabilities[6].

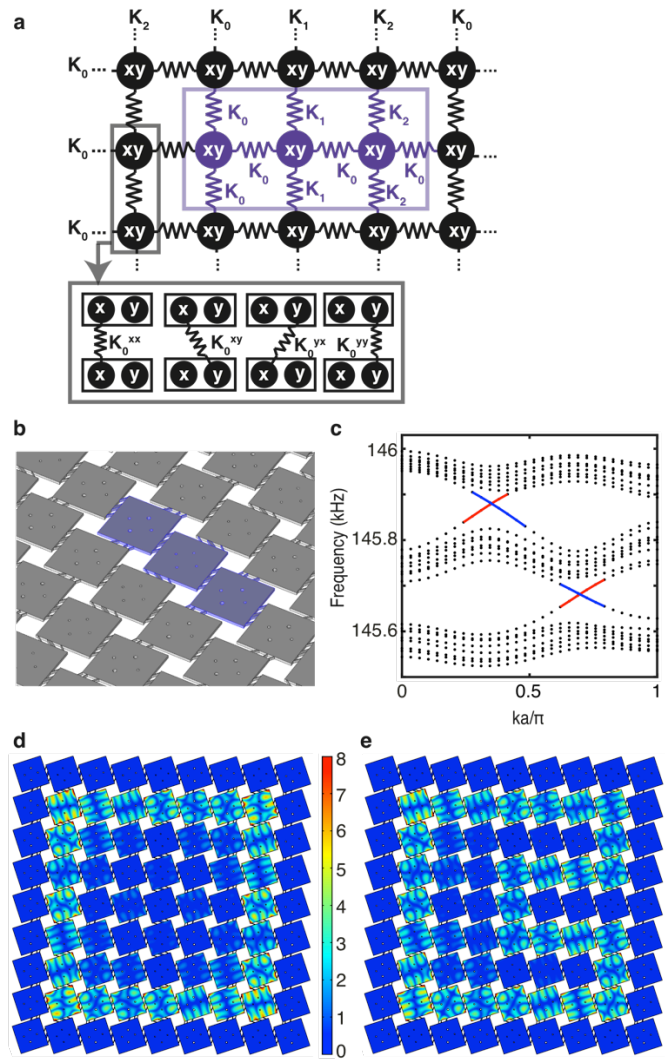

**Figure 4 | Topological insulator metamaterial. a.** Mass-spring model for topological insulator[15], where each mass x and y represents two degenerate modes of a plate in the metamaterial. Inset shows construction of the couplings between neighboring pairs of DOF *x* and *y*. **b.** Metamaterial

designed from topological insulator mass-spring model. **c.** Dispersion curves for the metamaterial, with periodicity in one direction and finite in the other, showing three bulk bands, indicated by the black points, separated by the two sets of counter-propagating edge modes, indicated by the red and blue solid lines. **d.** Example edge mode of the metamaterial. **e.** Edge mode propagation around a defect (3 fixed plates) in the metamaterial. In **d** and **e,** the outer edge plates are fixed, and the color bar indicates the amplitudes of both plots, in terms of total modal displacements with arbitrary normalized units.

**Methods**

The results presented in this paper have been calculated by using COMSOL Multiphysics® for the finite element simulations and MATLAB for the linear algebra calculations (except otherwise indicated). The two programs can communicate through the LiveLink® interface provided by COMSOL. The simulations were performed in ETH Euler cluster nodes and accessed up to 400 GB of RAM.

In all simulations we have used a linear elastic model with material parameters from epoxy resin for the beams (Young's modulus $E_B = 4.02$ GPa, Poisson ratio $\nu_B = 0.22$ and density $\rho_B = 1190 \ kg/m^3$) and steel for the plates ($E_P = 193$ GPa, $\nu_P = 0.3$ and $\rho_P = 8050$).

*Coupling Matrix Extraction*

We extract the coupling matrices by calculating the first 80 eigenmodes for a system of two plates coupled by beams (Fig 1c). The eigenmodes are calculated using COMSOL Multiphysics® 3D linear elasticity solver, with a highly refined mesh containing 964k elements. This fine mesh is required since the frequency shift introduced by the beams is much smaller than the resonance frequency of the plates, and therefore small imprecisions in the plate eigenfrequencies result in large errors in the calculated coupling matrix.

We then sample the x, y and z components of each eigenmode's displacement. The sampling is done at 2268 points at each plate, distributed over a test area extending 2 mm from the sides of the plate (Fig 1c). The sampled displacements for the 80 relevant eigenmodes are stored in a matrix, whose i-th column $U_j^i$ contains the *x, y* and *z* displacement for the i-th eigenmode: $U_j^i = \left(x_1^i, x_2^i, \ldots x_{2268}^i, y_1^i, y_2^i, \ldots, y_{2268}^i, z_1^i, z_2^i, \ldots, z_{2268}^i\right)^T$. The subindex *j* is used to distinguish between the two plates.

We express the displacement of the coupled-plate system in terms of a basis containing the first 40 normal modes of a free plate. Since our finite basis consists of a limited number of modes, it is incapable of exactly reproducing the coupled vibration profiles. For this reason, we use the Moore-Penrose pseudoinverse, which provides a least-square approximation to the solution. This approximation is given by $P_j = (A^T A)^{-1} A^T U_j$ where *A* is a matrix whose i-th column contains the displacement of the i-th free-plate eigenmode, sampled over the test area and organized in the same layout as $U_j$. We use 80 eigenmodes of the coupled system and 40 eigenmodes for the free plate. The mode selection must take into account several aspects: the number of eigenmodes for the two-plate coupled system should be twice the number of modes for the individual system, the coupled modes should not include any beam resonances, and families of degenerate modes should be either completely included or completely excluded. Once the matrix $P_j$ has been calculated, we assemble the matrix:

$$P = \begin{pmatrix} P_1 \\ P_2 \end{pmatrix},$$

and calculate the coupling matrix as $V = PDP^{-1} - H_0$ where $D$ is an 80x80 square matrix whose diagonal elements contain the eigenvalues of the coupled system, $D_{ij} = (2\pi f_i)^2 \delta_{ij}$ and $H_0$ contains the eigenvalues of an unperturbed single plate:

$$H_0 = \begin{pmatrix} D^0 & 0 \\ 0 & D^0 \end{pmatrix}$$

with $D_{ij}^0 = (2\pi f_i^0)^2 \delta_{ij}$ and $f_i^0$ the i-th eigenfrequency of an unperturbed plate.

The coupling matrix extraction method is equivalent to the first-order term of the Schrieffer-Wolff transformation, for a low-energy space spanning the first 40 eigenmodes of the unperturbed plate. This is because to first order, the Schrieffer-Wolff transformation is simply a restriction on the low-energy subspace, with the identity as a rotation matrix (see Supplementary Information). Once this first transformation has been performed, we calculate the higher orders to obtain the reduced-order model $V^R$ containing only the required modes (see supplementary information).

*Optimization Process*

<u>Combinatorial optimization</u>

We identify the optimal beam locations by performing an exhaustive search on combinations of beam locations and thicknesses. In this step, we calculate the coupling matrix $V_{ij}^R$ for a system containing multiple inter-plate coupling beams by adding together the coupling matrices of systems containing a single coupling beam. The validity of this approximation is examined in the Supplementary Information.

We first run the optimization code for different plate offsets in the range between 2 mm and 4 mm, with a spacing of 0.2 mm. We then assemble a table of coupling matrices $V_{ij}$ as a function of the beam location, for a fixed beam width of 0.2 mm. The coupling matrices are calculated using the finite element method described in the coupling matrix extraction section. The coupling matrices corresponding to beam widths other than 0.2 mm are calculated assuming a linear relation between beam width and coupling matrix, i.e. $V_{ij}(w) = [w/w_0]V_{ij}(w_0)$.

We then evaluate all combinations of beam locations (in steps of 0.1 mm) and beam widths (between 0.1 m and 0.5 mm in steps of 0.01 mm). We identify the optimal beam parameters by comparing the calculated coupling matrices to the objective coupling matrices. Before this comparison, the calculated and objective matrices $V_{ij}$ are normalized using the Hilbert-Schmidt norm $|V| = \sqrt{Tr[V^T V]}$. This is done since the exact norm can be adjusted after the fact by finely scaling the beam widths. This can be done due to the approximately linear relation between beam width and coupling matrix. In this step, we discard beam combinations whose norm is more than 50% off the target value, since those would result in extreme beam dimensions after rescaling.

The exhaustive search code is written in C++ to maximize its speed. For every offset (We considered 10 of them), the code explores $10^9 - 10^{10}$ configurations and takes between 43s and 197s to run on a 2.5 GHz Intel™ Core i7® laptop.

Gradient optimization

The gradient optimization is performed after the exhaustive search, in order to refine the beam parameters and account for interactions between the beams. At every gradient iteration, the coupling matrix is evaluated for a reference configuration and for small variations around this configuration. For systems containing three beams (the topological insulator and zero-dispersion material), the configuration is represented by a vector of the form $s = (x_1, w_1, \theta_1, x_2, w_2, \theta_2, x_3, w_3, \theta_3)^T$ where $x_i$ is the location of the i-th beam, $w_i$ is the i-th beam width and $\theta_i$ is the i-th beam angle. Different numbers of beams can be accommodated by adding or removing components. The coupling matrix $V_{ij}^R$ is also expressed in vector form $v = (V_{11}, V_{21}, V_{12}, V_{22})^T$. These definitions allow us to define a Jacobian matrix such as $v(s_0 + \Delta s) \approx v_0 + J\Delta s$ where the columns of $J$ are calculated as $J_i = v(s_0 + S_i) - v(s_0)$, and $S$ is a matrix where each column $S_i$ represents a perturbation in the configuration vector's i-th component. We use perturbations of 0.04 mm for the beam locations, 0.01 mm for the beam thickness and 2 degrees for the beam angles. The coupling matrix vector $v(s)$ is calculated using the coupling matrix extraction method described earlier.

The optimized state of the system after a gradient iteration is defined as $s_1 = s_0 + \alpha SJ^T(JJ^T)^{-1} * e - Sk$, where $e$ is the error vector $e = v - v_T$, $v_T$ is the objective coupling matrix $k$ expressed in vector form, $\alpha$ is a parameter controlling the gradient descent speed and $k$ is a vector from the Jacobian's kernel, i.e. $Jk = (0\ 0\ 0\ 0)^T$. The value of $\alpha$ is set to 0.4 at the beginning of the optimization process and then increased to 1 when the modulus of the error vector $e$ falls below 5% of the objective vector $v_T$'s modulus.

The kernel vector $k$ does not affect the coupling matrix and is chosen to minimize participation of unwanted modes. We observe that the addition of $k$ reduces these unwanted modes by 30% to 50%. We determine the direction of $k$ by first defining a scalar value $\epsilon$ that quantifies the participation of unwanted modes. The vector $k$ is then given by the projection of the gradient of $\epsilon$ into the kernel of $J$. The gradient of $\epsilon$ is defined with respect to the changes in the geometry, so $k = \gamma \wp_{Kern(J)} * \nabla_S \epsilon$, where $(\nabla_S \epsilon) = \epsilon(s_0 + S_i) - \epsilon(s_0)$, and $\wp_{Kern(J)}$ is a projector into the kernel of J. The value of $\epsilon$ is defined as $\epsilon = \sqrt{\epsilon_1^2 + \epsilon_2^2 + \cdots + \epsilon_n^2}$ where $n$ is the number of coupled modes within our frequency range of interest, and $\epsilon_i = |(I - P)U_i^{FEM}|$ is defined as the distance between the i-th coupled mode's displacement profile in the test area $U_i^{FEM} = (x_1^i, x_2^i, \ldots x_{2268}^i, y_1^i, y_2^i, \ldots, y_{2268}^i, z_1^i, z_2^i, \ldots, z_{2268}^i)^T$ and its projection into free-plate modes within the range of frequencies of interest, implemented with the projector $P = A(A^T A)^{-1} A^T$ where $A$ is a vector whose columns contain the sampled displacements of the free-plate modes in the frequency range of interest, following the same layout as $U_i^{FEM}$. The errors in the two coupled plates are reduced to a single number as $\epsilon$ by taking the RMS value of the two errors. The norm of $k$ is adjusted empirically between 0.5 and 2.

*Finite Element Simulations*

<u>Veselago lens</u>

Our model Veselago lens consists of 100x100 unit cells, each of them containing 141k elements. In order to solve this system, we follow a dynamic condensation approach[42]. We cut each unit cell at half along the length of the beams and define a transfer matrix that relates the displacements and forces acting on the boundary DOFS at the connection points, using 117 DOFs for every connecting beam cross section. We do this by first defining the unit cell dynamic force balance equation $-M\omega^2 \ddot{x} + ib\omega M x + V x = F$, where $M$ and $V$ are the unit cell's mass and stiffness matrices obtained from COMSOL, $b = 33\ s^{-2}$ is a damping parameter. We then introduce the dynamic stiffness matrix $D = -M\omega^2 + ib\omega M + V$ and decompose the set of nodal forces and displacements into set associated with boundary (b) and interior (i) nodes. By prescribing zero force at the interior nodes, the interior displacements can be condensate as $x_i = -D_{ii}^{-1}(D_{ib} x_b)$. As a result, we obtain a condensed matrix $D_{con} = D_{bb} - D_{bi}(D_{ii}^{-1} D_{ib})$. We solve this system of equations using the PARDISO solver included in the Intel Math Kernel Library, which can solve systems with multiple right hand sides without repeating common steps such as the matrix factorization. Similarly, we define a conversion matrix $C = \begin{bmatrix} -D_{ii}^{-1}D_{ib} \\ I \end{bmatrix}$, that provides the values of the full displacement vector $x$ as a function of the boundary DOF's $x_B$, $x = Cx_b$.

We then solve the force-balance problem for the full lens in terms of the boundary nodes. The force at each node is set to zero, except for those in the interface between the x=50 y=31 and x=50 y=32 which are driven with unit strength. After solving for the displacements in each step, we calculate the RMS amplitude of every unit cell by using the equation $x_{RMS} \propto \sqrt{E}$, where $E$ is the total steady-state energy stored in a unit cell, calculated as $E = (1/2)x_B^\dagger C^\dagger(\omega^2 M + V)Cx_B$ where † denotes the Hermitian conjugate.

Zero $c_g$ metamaterial and topological insulator

The zero group velocity metamaterial is simulated in COMSOL using a unit cell subject to Floquet boundary conditions at half the beam's length, using 1.02M elements per unit cell. The topological insulator dispersion relation (Fig. 4c), we model 4 unit cells (3x1 plates) stacked in a column in the finite dimension, such that the modeled system consists of 12x1 plates coupled with beams. This simulation was done using 901k elements. Fixed boundary conditions were applied on both ends of the beams of the finite dimension, and Floquet boundary conditions were applied in the other dimension. To determine the polarization of the edge modes around the crossing points, we calculate on which side of the model, in the finite dimension, the stored energy density is localized. The exact locations of the edge mode crossing points depend on which plate within the 3-plate unit cell is connected to the fixed boundary. The finite-size topological insulator was simulated directly in COMSOL using 2.6M elements.

## Acknowledgements

This work was partially supported by the ETH Postdoctoral Fellowship to K.H.M., and by the Swiss National Science Foundation. The authors would like to thank Osama Bilal for initial discussions.

## Author Contributions

K.H.M. & M.S. performed the optimization and simulation of the materials. K.H.M. wrote the manuscript, M.S. proposed the reduction approach and optimization algorithm. A.P. designed and performed the dynamic condensation. S.H. selected and interpreted the reduced order models. C.D. provided guidance during all stages of the project. All authors contributed to the discussion and interpretation of the results, and to the editing of the manuscript.

## Competing Financial Interests

The authors declare no competing financial interests.